\documentclass[12pt,preprint]{aastex}

\shorttitle{short title}
\shortauthors{Lin, Zhang \& Li}

\begin{document}

\title{Dominant gamma-ray bursts production in the early universe}

\author{J. R. Lin\altaffilmark{1}, S. N. Zhang\altaffilmark{1,2,3,4} and T. P. Li\altaffilmark{1,4}}
\affil{\altaffilmark{1}Tsinghua Center for Astrophysics, Tsinghua
University, Beijing, 100084, China}
\affil{\altaffilmark{2}Physics
Department, University of Alabama in Huntsville, Huntsville,
AL35899, USA}
\affil{\altaffilmark{3}Space Science Laboratory,
NASA Marshall Space Flight Center, SD50, Huntsville, AL35812, USA}
\affil{\altaffilmark{4}Institute of High Energy Physics, Chinese
Academy of Sciences, Beijing, China}

\bigskip

\begin{abstract}
It has been known that at least some of the observed gamma-ray bursts (GRBs)
are produced at cosmological distances and the GRB production rate may follow
the star formation rate. We model the BATSE detected intensity distribution
of long GRBs in order to determine their space density distribution and
opening angle distribution. Our main results are: the lower and upper distance limits to the GRB production are
$z\approx 0.24$ and $z>10$, respectively; the GRB opening angle follows exponential distribution
and the mean opening angle is about 0.03 radian; the peak luminosity appears a better standard candle than the
total energy of a GRB.
\end{abstract}


\maketitle

\section{Introduction}

Recent research results indicate that radiation from gamma-ray
bursts (GRBs) may be constrained mostly in narrow beams
\citep{wa98,fr99,wo01}, which suggests that GRBs may be standard
candles at cosmological distances \citep{fra01}. Previous studies
of the GRB space density distribution suggested that GRB
production rate may follow the observed star formation rate for
small values of redshift \citep{wij98}, but may increase
monotonically for very high values of redshift  \citep{sch,no02};
however neither the lower limit nor the upper limit of redshift
for GRB production could be determined reliably previously because
these studies are either limited by statistics or have to rely on
some empirical relationships between some statistical properties
of selected samples of GRBs. Because both the space density
distribution and beaming effect play important roles for the
observed GRB intensity distribution, here we make use of the
BATSE/CGRO GRB intensity distribution, which is the largest sample
of GRBs collected so far, to determine directly the GRB space
density and opening angle distribution.

\section{Models for GRB intensity distribution}

For the GRB space density distribution, we assume the following model, under the cosmological parameters of $\Omega=1$,
$H_0=50$ km s$^{-1}$ Mpc$^{-1}$ and $q_0=0.5$, in order to compare with previous star formation rate measurements \citep{st99}.

\begin{equation} n(1+z)=\left\{
\begin{array}{ll}
    0           ,   & \textrm{if}(z < z_0)\\
  n_{0}(1+z)^{3.5}, & \textrm{if}(z_0\leq z<z_{break}) \\
 n_{0}(1+z_{break})^{3.5}, & \textrm{if}(z\geq z_{break})
\end{array}\right..
\end{equation}
The functional form is taken from the measured star formation rate \citep{st99} with $z_0>0.2$ and $z_{break}\approx 1$.
Our main goal in this paper is to determine
$z_0$ and $z_{break}$ for GRBs from the observed GRB intensity distribution, in order to answer the question
where in the universe GRBs are produced predominantly.

For the GRB opening angle distribution, we study three cases when the GRB opening angle $\theta$
follows Gaussian, exponential, or power-law distributions. For a Gaussian
distribution of $\theta$, the probability that the opening angle equals to a
given angle is,

\begin{equation}
P(\theta)=\frac{1}{\sqrt{2\pi}\sigma}\textrm{e}^{-\frac{(\theta -
\theta_{0})^{2}}{2\sigma^{2}}}.
\end{equation}
We assume that the total energy $E_0$ is the same for for every
long GRB, i.e., they can be used as standard
candles\citep{fra01,bal03} and the opening angle $\theta$ is very
small, the probability that the observed fluence $f$ is higher
than a specified value $F$ is then,
$$ P(f\geq
F)=P(\Omega \approx \frac{\pi}{4}\theta^{2}
\leq\frac{E_{0}(1+z)}{Fd_L^{2}})$$
\begin{equation}
=\int_{0}^{(\frac{4E_0(1+z)}{\pi
Fd_L^{2}})^{\frac{1}{2}}}\frac{\theta^{2}}{16\sqrt{2\pi}\sigma}\textrm{e}^{-\frac{(\theta
-\theta_{0})^{2}}{2\sigma^{2}}}\textrm{d}\theta,
\end{equation}
where $d_{L}$ is the luminosity distance. Thus the number of
bursts detected is,
\begin{equation}
N=\int \frac{1}{4}\pi n
d_L^{2}\textrm{d}d_L\int_{0}^{(\frac{4E_0(1+z)}{\pi
Fd_L^{2}})^{\frac{1}{2}}}
\frac{\theta^{2}}{\sqrt{2\pi}\sigma}\textrm{e}^{-\frac{(\theta -
\theta_{0})^{2}}{2\sigma^{2}}}\textrm{d}\theta.
\end{equation}

Similarly, for an exponential distribution with the parameter
$\lambda$,
\begin{equation}
N=\int \frac{1}{4}\pi n
d_L^{2}\textrm{d}d_L\int_{0}^{(\frac{4E_0(1+z)}{\pi
Fd_L^{2}})^{\frac{1}{2}}} \theta^{2}\lambda\textrm{e}^{-\lambda
\theta }\textrm{d}\theta,
\end{equation}
and for a power-law distribution with the parameter $t$,
\begin{equation}
N=\int \frac{1}{4}\pi n
d_L^{2}\textrm{d}d_L\int_{0}^{(\frac{4E_0(1+z)}{\pi
Fd_L^{2}})^{\frac{1}{2}}} \theta^{2-t}\textrm{d}\theta.
\end{equation}

If we assume the peak luminosity of GRBs is the standard candle, we only need to replace $E_0$ and $F$ by the standard candle luminosity $L_0$ and
the peak flux $P$ in  equations (3), (4), (5) and (6), respectively. We number these slightly modified equations as (3)$'$,
(4)$'$, (5)$'$ and (6)$'$, respectively;
for brevity we do not list these equations explicitly.

\section{Results}

We apply equation (1) for the GRB
space density distribution and equations (4), (4)$'$, (5), (5)$'$, (6) and (6)$'$ for the GRB opening angle distribution to fit the observed GRB
intensity distribution. We focus our study on long GRBs ($T_{90} > 2 s$) \citep{kou93} because currently only long bursts are known to
be originated at cosmological distances from the direct redshift measurements of the optical afterglows of some
long GRBs. All long GRBs with peak flux values and detection probability above 5\% included in the GRB 4B+ Catalog
 \citep{pa00} are selected, resulting in 1727 GRBs as our sample. The fluence values of these GRBs are taken also directly
 from the catalog. For the peak flux distribution, the detection efficiency correction is done directly using the detection
 probability of each peak flux value given in the catalog.
 Because the fluence and peak flux does not have a good linear correlation, for each fluence range we average
 the detection probability
 of all GRBs included in this fluence range and then correct for the detection efficiency of this fluence range.

\clearpage

\begin{table}
\begin{center}
Total energy as standard candle.
\begin{tabular}{c@{}|c@{}c@{ }c@{ }c@{}} \hline
&\textbf{Gaussian}&\textbf{exponential}&\textbf{power-law}\\
\hline
$\chi^{2}_r$             &$281.3/194$               &\textbf{261.1}/\textbf{195}             &$623.2/195$& \\
\hline
                         &$\theta_{0}$=0.028$\pm$0.007\\
\raisebox{1.2ex}[0pt]{}&$\sigma$=0.050$\pm$0.008    &\raisebox{1.2ex}[0pt]{$\lambda$=\textbf{38.1}$\pm$\textbf{3.8}}&\raisebox{1.2ex}[0pt]{$t$=4.51$\pm$0.09}\\
\hline
$z_{0}$                  &$0.23\pm0.07$             &\textbf{0.23}$\pm$\textbf{0.08}         &$0.27\pm0.11$\\
$z_{break}$              &$7.7-\infty$              &\textbf{7.9-$\infty$}                   &$7.2-\infty$\\
\hline
$n_{0}$                  &$231\pm51$                &$215\pm45$                              &$207\pm44$\\
$E_{0}$                  &$0.69\pm0.37$             &$0.80\pm0.41$                           &$0.72\pm0.36$ \\
\hline
\end{tabular}

Peak luminosity as standard candle.
\begin{tabular}{c@{}|c@{}c@{ }c@{ }c@{ }} \hline
&\textbf{Gaussian}&\textbf{exponential}&\textbf{power-law}\\
\hline
$\chi^{2}_r$             &$241.1/194$               &\textbf{227.1/195}             &$539.7/195$ \\
\hline
                         &$\theta_{0}$=0.031$\pm$0.008\\
\raisebox{1.2ex}[0pt]{}&$\sigma$=0.047$\pm$0.01    &\raisebox{1.2ex}[0pt]{$\lambda$=\textbf{34.4}$\pm$\textbf{2.1}}&\raisebox{1.2ex}[0pt]{$t$=4.52$\pm$0.09}\\
\hline
$z_{0}$                  &$0.23\pm0.09$             &\textbf{0.24}$\pm$\textbf{0.07}         &$0.29\pm0.09$\\
$z_{break}$              &$9.5-\infty$              &\textbf{9.8-$\infty$}                   &$9.2-\infty$\\
\hline
$n_{0}$                  &$273\pm44$                &$264\pm49$                              &$257\pm39$\\
$L_{0}$                  &$0.67\pm0.32$             &$0.64\pm0.22$                           &$0.69\pm0.30$ \\
\hline
\end{tabular}
\end{center}
\caption{$\chi^{2}_{r}=\chi^{2}/${\it dof}. $L_0$ is in units of $10^{51}$ {\it
ergs/s}. $n_0$ in units of Gpc$^3/$year. All errors quoted are for 68.3\%
confidence. For both cases the exponential distribution of the opening angle offers the best fits; however
the residuals for the combination of peak luminosity as the standard candle and the exponential distribution is the only
statistically acceptable one.}
\end{table}

\clearpage


\begin{figure}[]
\centerline{
\epsscale{0.7}
\plotone{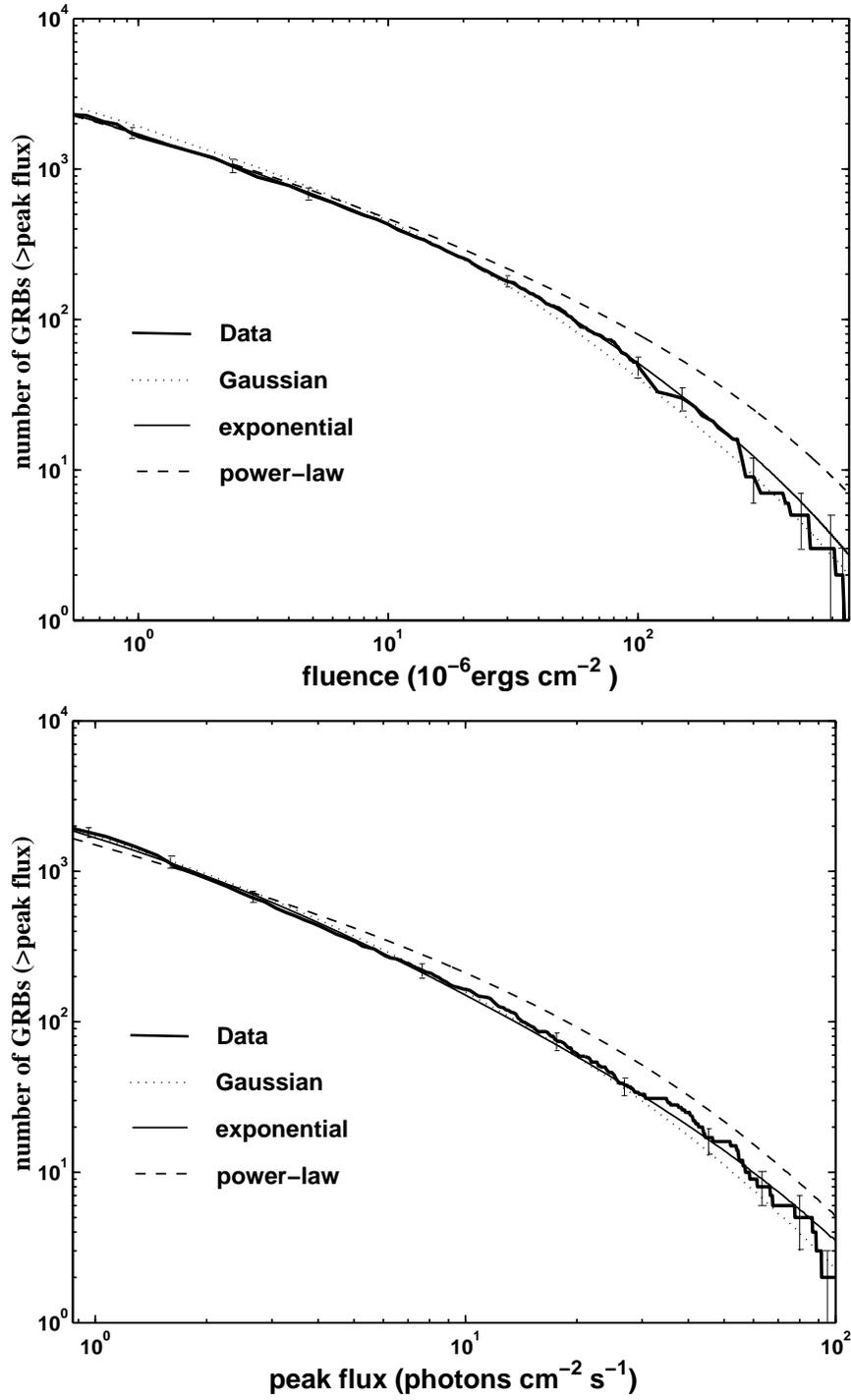}} \caption {Comparison of different
models with data. Top panel: long GRBs with fluence as standard
candle; Lower panel: long GRBs with peak luminosity as standard
candle.}
\end{figure}


\clearpage

Our fitting
results are shown in $\emph{Table.1}$ and $\emph{fig.1}$. Because our fittings are done with the integral distributions, the
numbers of GRBs in different bins are not completely independent. To address this problem, we carried out extensive bootstrap
simulations. Taking the 1727 GRBs as the seed distribution, we randomly generated a large numbers of peak flux or fluence distributions
and fitted the sampled distributions in the same way. The best values and errors of all parameters are then calculated
from these fittings; the bootstrapped parameter values and errors are consistent with that listed in Table 1. We therefore believe
the correction between the different points does not have any significant impact to our results.
From the fitting results, we conclude:
\begin{itemize}

\item For long GRBs, despite the different fitting residuals of the two standard candle models and three opening angle distributions
assumed, the cosmological parameters of GRBs inferred are strikingly similar, indicating the robustness of the GRB cosmological
model.

\item The power-law distribution for the GRB opening angles cannot fit the data and is thus rejected confidently.

\item The Gaussian distributions for the GRB opening angles fit the data only marginally, and may be rejected too.

\item For both standard candle models the exponential distribution for the GRB opening angles provides the best fit to the data.

\item The combination of the peak luminosity as standard candle and the exponential distribution for the GRB opening angles
is the only one with statistically acceptable fit to the data; we therefore accept this as the accepted model for long GRBs. Under
this model the peak luminosity of a GRB is $(0.64\pm0.22)\times10^{51}$ {\it ergs/s}, the exponential constant of the
opening angle distribution is $34\pm2$ (radian$^{-1}$), the lower bound to the GRB redshift is $z_0=0.24\pm0.07$, and the
upper break redshift is $z_{break}>9.8$ (no lower limit to the break is detected because not enough sufficiently faint GRBs have been
observed with BATSE).

\end{itemize}

\section{Discussion}
First we compare our study with previous studies of GRB cosmological models with BATSE intensity distribution
\citep{fen93,fen95}\\ \citep{wic93,wij98}. All these studies concluded that the dimmest GRBs are at redshift around unity, probaly
following the measured star formation rate which peaks also at redshift around unity \citep{st99}. The main differences
between our approach and previous studies are the following:

\begin{itemize}
\item We take the GRB opening angle distribution as part of the
cosmological model. This is very important, since the observed GRB flux depends strongly on the opening angle, as well as
the redshift.

\item We separate the long and short GRBs as two different classes in our study. Our conclusions are only applicable to
long GRBs. For short GRBs, no conclusion can be drawn due to the limited statistical quality of data.

\item One simplification of our study
is that we did not make the energy spectral correction to the fluence or peak flux as a function of $z$ (K-correction); we just take
the fluence or peak flux values as listed in the catalog. Technically this correction
is not straightforward, because we can no longer uniquely convert the fluence or peak flux directly to redshift, because we believe
that GRB radiation is beamed, and the opening angle follows a certain distribution. Therefore the observed fluence or peak flux
is due to both the redshift and the opening angle of the GRB. The good fit of the assumed simple cosmological model with only
several free-parameters indicates that this correction may not impact our conclusions significantly. In fact, since the
energy spectra of GRBs in the BATSE band follows approximately a power-law with an index of $-$2, the K-correction is not important.

\end{itemize}

Frail {\it et al.} (2001) suggested that the total energy of GRBs may be a constant, based on the opening angles
of several GRBs determined from their afterglow light curve break. Our result is not in major conflict with their suggestion, but favors the
peak luminosity as the standard candle. Perhaps the direction of the GRB beam is not completely stable, as indicated from
observations of other relativistic jets and outflows from stellar mass and supermassive black holes; only when the beam
is pointed directly at us can we observe the maximum flux from the GRB. However it has been argued that for some GRBs
the lower limit of the total energy released is $10^{52}$ {\it ergs} as determined from the redshifted X-ray emission lines \citep{Ghi02}. However
the reality of the detection of the emission lines is still under debate, and an alternative
interpretation of the detected emission lines also exists \citep{cer}.

\clearpage

\begin{figure}\label{fig:fig2}
\centerline{
\epsscale{0.7}
\plotone{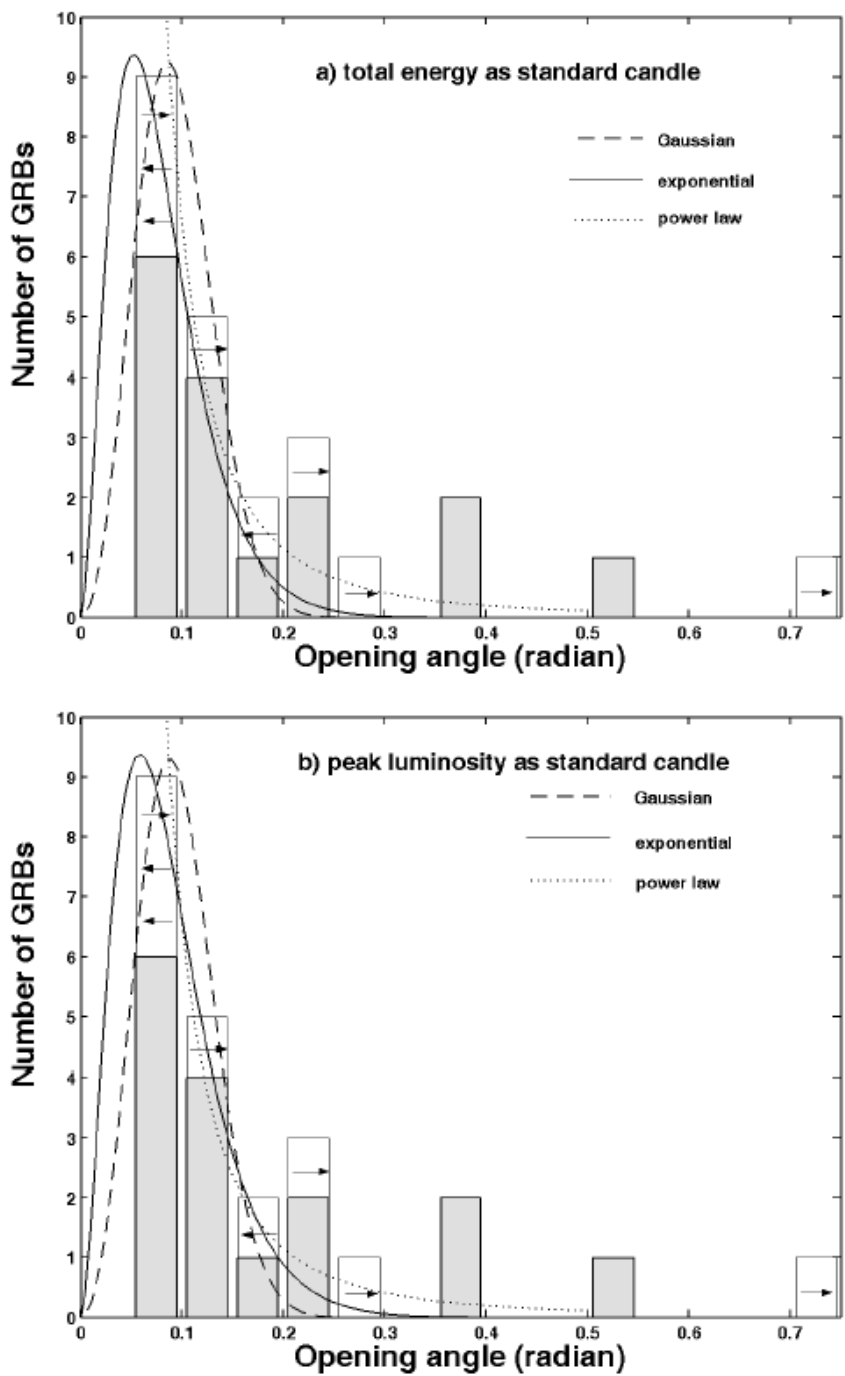} } \caption {The data for the obsered
GRB opening angles are taken from Bloom  {\it et al.} 2003; arrows
indicate lower and upper limits of GRB opening angles. The
power-law distribution fits the data for large opening angles, but
fails to predict the absence of GRBs detected with opening angles
smaller than 0.04 radian. The Gaussian distribution agrees with
the peak reasonably well, but cannot predict enough GRBs with
opening angles greater than 0.2 radian. The exponential
distribution is consistent with the data.}
\end{figure}

\clearpage

The opening angles of some GRBs with afterglow detections have
been determined previously from the GRB afterglow data
\citep{fra01,bloom2003}; we call this Frail-distribution. In
$\emph{fig.2}$ we compare the Frail-distribution with the opening
angle distributions determined with the observed GRB intensity
distribution data, just as a sanity check. The power-law
distribution agrees with the Frail-distribution for large opening
angles, but fails to predict the absence of GRBs detected with
opening angles smaller than 0.04 radian. No GRBs with opening
angles narrower than 0.04 radian was directly observed, but the
recent report on the polarization of GRB021206 suggests that it is
possible for some GRBs to have opening angles narrower than 0.01
radian \citep{cob03,wax03}. The Gaussian distribution agrees with
the peak of the Frail-distribution distribution reasonably well;
however this model cannot predict enough GRBs with opening angles
greater than 0.2 radian. The exponential distribution is
consistent with the Frail-distribution distribution for both
standard candle models, but favors slightly the peak luminosity
standard candle model. It was suggested that for the model in
which the GRBs follow the star formation rate strictly, the most
possible opening angle should be around 0.1 radian
\cite{bloom2003}. However, if the majority of faint GRBs originate
at higher redshift than the peak of star formation rate as our
model predicts, then a narrower average opening angle is required.

We therefore
conclude that GRB opening angles follow an
exponential distribution with a mean opening angle of $1/\lambda=0.03$ radian,
and the GRB space density follows a power-law with an index of $-$3.5 between $z_0=0.24\pm0.07$ and $z_{break}>10$.
The general trend
of the GRB space density distribution determined here as a function of $z$ is consistent with previous studies \citep{sch,no02}; however
our result is statistically much more robust, and in particular $z_0$ is determined for the first time and the lower limit
to $z_{break}$ is constrained more strictly than before. Our
result therefore demonstrates that
GRBs are no longer produced frequently in the nearby and present universe at $z<0.24\pm0.07$, and GRBs are predominantly
produced in the early universe at $z>10$. Indeed most GRBs with direct redshift measurements
have redshifts $z>0.3$ \citep{djo01}, with the exceptions of
GRB 980425 ($z$=0.0085) and GRB 030329 ($z$=0.1687) which may be associated with two peculiar but similar supernovae SN1998bw \citep{sn1998},
and SN2003dh \citep{sn03}. It remains to be seen if these GRBs belong to a subset of GRBs which are different from
the majority of GRBs which seem to be produced at much larger redshift.

It has been suggested that GRBs follow star formation rate, because it is commonly believed that GRBs are produced
as the final gravitational collapse of massive stars \citep{djo01,pac98}. The GRB space density determined here follows the measured star formation
rate between $z\approx 0.24$ to $z\approx 1$. The determination of $z_0=0.24\pm0.07$
suggests that the massive stars capable of producing the majority of long GRBs are no longer formed frequently today.
Future GRB experiments capable
of measuring the redshift of many more GRBs and thus providing direct distance measurements of GRBs,
such as the {\it Swift} mission to be launched in late 2003 \citep{swift},
will test this prediction firmly.

The monotonic increase of the GRB space density up to at least
$z=10$ suggests that the formation rate of massive stars capable
of producing GRBs should peak at significantly higher $z$ than the
measured star formation rate. This implies that the ``dark ages"
of the universe ended much earlier than previously believed at
$z\sim 4-5$ based on the measured space density distribution of
high-redshift QSOs \citep{hook}. There are also some previous
independent works showing that an essential amount of long GRBs
should be at $z > 4$ \cite{cho02,bag03} or even up to $z \cong
20$\cite{mes96}. A recently detected GRB031203 by INTEGRAL was
reported to be located at $z > 9$ (formally $z \geq 11$)
\cite{grb03}, consistent with our model prediction. Therefore GRBs
may be a probe of the epoch of reionization of the early universe
\citep{lamb03}. Currently the GRB intensity distribution data
cannot constrain the exact upper limit of $z_{break}$, because not
enough extremely faint GRBs have been observed. Future GRB
experiments, such as the proposed ``Next Generation GRB Mission"
\citep{exist} would detect many more GRBs originated in the
extremely early universe, which may determine the exact ending
time of the ``dark ages" of the universe in order to test the
prediction of some numerical simulations \citep{first-star} which
suggests that the first stars may have been formed at $z\sim 20$.
GRBs may prove to be the most sensitive and ultimate probes into
the early epoch of the universe when the first structures were
formed.

\noindent {\bf Acknowledgement: } We thank Drs. Zigao Dai, Tan Lu, Yongfeng Huang and Daming Wei for valuable comments to the manuscript.
This study
is supported in part by the Special Funds for Major State Basic Research Projects (10233010) and by the National Natural Science Foundation of China.
SNZ also acknowledges supports
by NASA's Marshall Space Flight Center and through NASA's Long Term Space Astrophysics Program.

\medskip
\noindent {\small Correspondence should be addressed to S.N.Z. (e-mail: zhangsn@tsinghua.edu.cn).}
\end{document}